\documentclass[a4paper,fleqn,usenatbib]{mnras}

\usepackage{newtxtext,newtxmath}

\usepackage[T1]{fontenc}
\usepackage{ae,aecompl}

\usepackage{graphicx}	
\usepackage{amsmath}	
\usepackage{amssymb}	

\usepackage{color}
\usepackage[usenames,dvipsnames,svgnames,table]{xcolor}
\usepackage{subfigure}
\usepackage{multirow}
\usepackage{float}
\usepackage{nameref}
\usepackage{lineno}

\title[Diffuse $\gamma$-ray emission from self-confined CRs]{Diffuse $\gamma$-ray emission from self-confined cosmic rays around Galactic sources}

\author[]{
	M. D'Angelo $^{1}$\thanks{E-mail: marta.dangelo@gssi.infn.it}
	G. Morlino $^{1,2}$\thanks{E-mail: giovanni.morlino@gssi.infn.it}
	E. Amato $^{2}$\thanks{E-mail: amato@arcetri.astro.it}
	P. Blasi$^{2,1}$\thanks{E-mail: blasi@arcetri.astro.it}
	\\
	$^{1}$ Gran Sasso Science Institute, Viale F. Crispi 7 - 67100 L' Aquila, Italy\\
	$^{2}$ INAF/Osservatorio Astrofisico di Arcetri, Largo E. Fermi, 5 - 50125 Firenze, Italy
}

\date{Accepted 2017 October 30. Received 2017 October 6; in original form 2017 June 29}

\pubyear{2017}

\begin{document}
\label{firstpage}
\pagerange{\pageref{firstpage}--\pageref{lastpage}}
\maketitle

\begin{abstract}
The propagation of particles accelerated at supernova remnant shocks and escaping the parent remnants is likely to proceed in a strongly non-linear regime, due to the efficient self-generation of Alfv\'en waves excited through streaming instability near the sources. Depending on the amount of neutral hydrogen present in the regions around the sites of supernova explosions, cosmic rays may accumulate an appreciable grammage in the same regions and get self-confined for non-negligible times, which in turn results in an enhanced rate of production of secondaries. Here we calculate the contribution to the diffuse gamma-ray background due to the overlap along lines of sight of several of these extended halos as due to pion production induced by self-confined cosmic rays. We find that if the density of neutrals is low, the halos can account for a substantial fraction of the diffuse emission observed by Fermi-LAT, depending on the orientation of the line of sight with respect to the direction of the Galactic centre.
\end{abstract}




\section{Introduction}
\label{sec:Intro}

The theory of the origin of Galactic cosmic rays (CRs) is based on two pillars: 1) CRs are injected from sources, perhaps supernova remnants, with an approximately power law spectrum, as would be warranted by diffusive shock acceleration. 2) CRs propagate in the Galaxy following a combination of diffusive motion in magnetic disturbances and advection. These two aspects of the problem and their observational signatures are discussed in detail in some recent review papers \cite[]{blasi2013,amato2014}. 

The main information about the transport of CRs in the Galaxy comes from the investigation of the so called cosmic clocks: on one hand CR nuclei propagating throughout the Galaxy and occasionally interacting with gas in the disc produce lighter stable nuclei through spallation, and their equilibrium abundance as measured at the Earth location provides an estimate of the grammage traversed by CRs. On the other hand, unstable nuclei produced in the same spallation reactions tend to disappear by radioactive decays, so that the ratio of abundances of unstable and stable isotopes returns an estimate of the confinement time in the Galactic halo. 

While the B/C ratio is now measured up to energies of $\sim$ TeV per nucleon, and hence provides information about grammage in the same energy range, the flux of unstable isotopes, such as $^{10}$Be, has been measured only up to energies of the order of $\sim$GeV/n \citep{Yanasak2001}, so that the confinement time can only be constrained in such a low energy region. This leaves room for speculations about the fact that at least some of the grammage that CRs traverse is actually accumulated near their sources. This has been discussed in alternative models of CR transport, such as the nested leaky box model \cite[]{2010PhRvD..82b3009C} (see also \cite{2017PhRvD..95f3009L}), as well as in connection with the possibility to measure enhanced gamma ray emission from the near source regions \cite[]{1996A&A...309..917A} if the diffusion coefficient near the sources happens to be much smaller than on average.  

Recently, the non-linear effects induced by the large density and large density gradients in the near source regions have been studied in detail and found to be potentially responsible for a substantial decrease in the diffusion coefficient as due to self-generated waves \cite[]{ptuskin2008,Malkov2013,Nava2016,dangelo2016}.

Some investigations of the non-linear problem in which CRs affect their own diffusion have been presented by \cite{ptuskin2008} and \cite{Malkov2013} by using self-similar solutions of the non-linear transport equation. Scenarios involving non-linear CR diffusion on Galactic scales have also been recently discussed by \cite{Blasi-Amato-Serpico2012,2013JCAP...07..001A,2015A&A...583A..95A}. The time dependent transport equation in the presence of self-generated waves and damping near sources of CRs was recently solved numerically by \cite{Nava2016} and \cite{dangelo2016}. The latter aimed at calculating the grammage accumulated by CRs in the near source region. This investigation found that the resulting grammage heavily depends on the abundance of neutral hydrogen in the vicinity of the sources, in that it determines the level of ion-neutral damping for the self-generated waves. In the absence of neutrals the phenomenon of self-confinement can account for a substantial part of the grammage usually attributed to propagation on galactic scales. 

In this paper we investigate a byproduct of the phenomenon of self-confinement of CRs in the near source regions, namely the production of gamma radiation due to pion production by CRs leaving the source and interacting with ambient gas. This process results in the production of regions of extended gamma ray emission around locations where supernovae exploded. While the detection of a single  halo is prohibitive with current gamma-ray telescopes, the superposition of the gamma-ray emission from many such halos along the line of sight induces a contribution to the diffuse gamma-ray background in the Galactic disc region. Here we discuss the extent to which this contribution may be important for and/or detectable through observations, by comparing our results with the findings of \cite{fermi2016} and \cite{yang2016} based on \emph{Fermi}-LAT measurements.

In \S\ref{sec:CR_dist} we briefly illustrate how the CR spectrum escaping an individual source is computed. The technical aspects of this calculation were already described in detail by \cite{dangelo2016}. In \S\ref{sec:gamma} we calculate the $\gamma$-ray emission produced by these escaping particles through nuclear collisions. In \S\ref{sec:tot_flux} we describe our procedure for sampling, in time and space, the SNR distribution in the Galactic disc. Finally in \S\ref{sec:results} we sum up the estimated contribution of all the relevant SNRs in given portions of the sky and compare our results with available data. A critical discussion of limitations and implications of our results is presented in \S\ref{sec:conclude}.

\section{CR confinement around SNRs}
\label{sec:CR_dist}

In this section we summarize the findings of recent work by \cite{dangelo2016} concerning particle propagation regulated by self-generated turbulence, namely by the perturbations to the regular magnetic field induced by the escaping particles themselves. In mathematical terms, this task requires the solution of a system of two coupled time-dependent equations, one describing the particle transport and the other one describing the associated development of turbulence. 

Let us consider a supernova exploding in the disk of our Galaxy and focus on the escape of particles accelerated in its aftermath. 
The magnetic field of the Galaxy in the disc region can be considered as the overlap of two components, a large scale ordered field with strength $B_{0}$ and a turbulent component that we can think of as being described by a strength $\delta B$ and a power spectrum as a function of spatial scale. For well behaved spectra the power is concentrated around a large scale $\lambda$ that is the energy containing scale. If $\delta B/B_{0}\sim 1$, on a scale $\sim \lambda$ the magnetic field can be considered, in first approximation, as roughly ordered while the magnetic field line changes direction by order unity on larger scales. If $\delta B/B_{0}\ll 1$ the change of direction actually takes place on scales much larger than $\lambda$. In a recent paper by \cite{beck2016} the authors carry out a complex analysis of the intensity and polarization of the Galactic synchrotron emission to constrain the coherence scale of the disordered component of the magnetic field: they conclude that $\lambda\sim 220$ pc while a lower limit of $50$ pc (with $5\sigma$ confidence level) can be also imposed. This scale appears to be appreciably larger than the typical size of a SNR in its period of highest efficiency of CR acceleration. Hence, also following previous suggestions by \cite{ptuskin2008}, we describe particle propagation outside the remnant on scales smaller than $\lambda$ as one-dimensional \cite[see also][]{dangelo2016,Malkov2013,Nava2016}). We setup the problem with a source located at the centre of a region with size $\lambda=2L_{c}$ with $L_{c}\simeq 100$ pc. In realistic cases we do expect a complex transition between the near-source region, where the propagation can be assumed as 1D, and the far-source region where diffusion becomes 3D \cite[see, e.g. simulation by][]{Giacinti2013}. Here we do not aim at describing such a complex transition region but we assume that at a distance $L_c$ the CR spectrum reduces to the average Galactic one.
The CR density and density gradients in the region around the source are large enough to drive a streaming instability that in turn produces waves that can scatter particles, thereby confining them in the near-source region for long times. 

Before energetic particles start affecting the environment, the turbulence level is assumed to be the same as the Galactic average and the diffusion coefficient, $D_g$, is assumed to correspond to the one estimated by fitting the CR data with GALPROP~\citep[][see also \texttt{http://galprop.stanford.edu}]{galprop1998}. Assuming a Kolmogorov's turbulence spectrum, we adopt for $D_g$ the analytical expression derived by ~\cite{ptuskin2009} as a fit to GALPROP results within a leaky box model: $D_g(E) = 3.6 \times 10^{28} E_{\mathrm{GeV}}^{1/3} \, \mathrm{cm^2 s^{-1}}$. The 1D approximation is only valid to describe the propagation of particles up to a distance $|z|<L_c$, and the obvious requirement is that the particle mean free path be less than $L_c$. In mathematical terms this condition is written as $3 D_g(E)/c \ll L_c$, and is satisfied for particle energies up to $10^6 \, \mathrm{GeV}$. In the following we restrict our analysis to CRs with energy in the range $ 10 \, \mathrm{GeV} \leq E \leq 10^4 \, \mathrm{GeV}$.

The evolution of the CR distribution function in the source vicinity is determined by advection, diffusion and adiabatic expansion (or compression). Within the flux tube approximation the $1$D CR transport equation reads
\begin{equation}
\frac{\partial f}{\partial t} + u \frac{\partial f}{\partial z} - \frac{\partial }{\partial z} \left[ D(p,z,t) \frac{\partial f}{\partial z} \right] - \frac{du}{dz} \frac{p}{3}\frac{\partial f}{\partial p}= 0 \, .
\label{eq:transp}
\end{equation}
Equation~\eqref{eq:transp} does not contains any source term because the injection of particles is mimicked using an appropriate initial condition (see below).

The velocity appearing in the advection term, $u$, is $u = +v_A$ for $z>0$ and $u = -v_A$ for $z<0$, where $v_A = B_0/\sqrt{4 \pi n_i m_i}$ is the Alfv\'en speed, namely the characteristic propagation velocity of magnetic disturbances in a plasma, here defined in terms of the unperturbed magnetic field $B_0$ and of the ISM ions mass ($m_i$) and density ($n_i$). The spatial derivative of $u$ appearing in the adiabatic term will then read: $du/dz = 2v_A \, \delta(z)$. 

In equation~(\ref{eq:transp}) energy losses have been neglected since the time scale for $pp$ scattering is of order $t_{\rm loss} \sim 7 \times 10^7 (n_{\rm gas}/0.45 \, \rm cm^{-3})^{-1} \mathrm{yr}$, where $n_{\rm gas}$ is the gas density in the disc of the Galaxy. Such time is much longer than the typical confinement times that will result from our calculations (as checked {\it a posteriori}). Since we restrict our attention to energies above $\sim 10$ GeV, other losses, such as ionization, can be neglected as well.

At $t=0$, the population of particles of given momentum $p$ comprises diffuse Galactic CRs, described by the distribution function $f_g(p)$ and particles escaping the source and making their way into the ISM. In terms of spatial distribution, the former are taken as uniform throughout the flux tube, while the latter are assumed to form a gaussian of half width $z_0\ll L_c$, centred in $z=0$, so that the total initial particle distribution function reads:
\begin{equation}
f(p,z,t=0) = q_0(p) \exp\left[-\left ( \frac{z}{z_0} \right)^2\right] + f_g(p) \, .
\label{eq:ic}
\end{equation}
The assumption that at $t=0$ the freshly accelerated particles are distributed as a gaussian is made for purely numerical reasons: in the presence of self-generation of waves, at very early times, the density of particles close to the source is so large that the local diffusion coefficient becomes extremely small, thereby making numerical convergence rather problematic. Of course this phase lasts very shortly so that its effect on the total grammage accumulated by particles during their escape from the near source region is negligible. We checked this by adopting different values of the width of the gaussian, $z_{0}$, and testing that the physical results do not change in any appreciable way.

The galactic proton spectrum, $f_g$, has been taken to equal that resulting from the most recent AMS-02 measurements~\citep{AMS02proton}:
\begin{equation}
f_g(p) = 6.8 \times 10^{22} \left(\frac{p}{45 p_0}\right)^{-4.85} \left[1 + \left ( \frac{p}{336 p_0} \right )^{5.54}\right]^{0.024} \mathrm{\left(\frac{erg}{c}\right)^{-3}} \mathrm{cm^{-3}} \, ,
\label{eq:fg}
\end{equation}
where $p_0 = m_p c$, with $m_p$ the proton mass and $c$ the speed of light.

As for fresh particles, their initial spectrum is $q_0(p) = A (p/p_0)^{-\alpha}$, with $\alpha=4$ as appropriate for acceleration at a strong shock wave (but we will also analyze a steeper spectrum with $\alpha=4.2$). The normalization constant, $A$, is related to the SNR energetics by imposing that a fraction $\xi_{CR}$ of the kinetic energy of the SNR is converted into accelerated particles:
\begin{equation}
A = \frac{\xi_{CR} E_{SN}}{\pi R_{SN}^2 \mathcal{I}} \, ,
\end{equation}
where
\begin{equation}
\mathcal{I} = \int_{-L_c}^{L_c} dz \exp\left[-\left ( \frac{z}{z_0} \right)^2\right]\int_{0}^{\infty} dp 4\pi p^2 \left(\frac{p}{p_0}\right)^{-\alpha}\epsilon(p) \, .
\end{equation}  
In the above expressions $E_{SN}$ is the kinetic energy of the SNR and $R_{SN}$ its radius, while $\epsilon(p)$ is the kinetic energy of a particle of momentum $p$. In the following the SNR kinetic energy is taken to be $E_{SN}=10^{51}\, \mathrm{erg}$, and $\xi_{CR}=20\%$. For the SNR radius we adopt a value $R_{SN}\approx 20$ pc, corresponding to the slowly varying size of the SNR during the Sedov phase. Finally, for $z_0$ we take $z_0=1 \, \mathrm{pc}$, which ensures the condition $z_0\ll L_c$, while still allowing reasonable computation times. We have checked that this choice of $z_0$ does not affect the results.

The symmetry of the problem with respect to $z=0$ allows us to solve the equations in the half-space $0 \le z \le L_c$. 
Equation~(\ref{eq:transp}) is a second order partial differential equation and two boundary conditions are needed to determine a particular solution. The first boundary condition we impose comes from the integration of equation~\eqref{eq:transp} around $z=0$, which leads to:
\begin{equation}
\left . D(p, z=0, t) \frac{\partial f}{\partial z}   \right |_{z=0} = - \frac{v_A}{3} p \left . \frac{\partial f}{\partial p}  \right |_{z=0}  \, .
\label{eq:bc1}
\end{equation}
The second boundary condition comes from imposing that the distribution function be equal to the Galactic CR spectrum $f_g(p)$ at $|z|=L_c$:  
\begin{equation}
f(p,z=L_c,t) = f_g(p)\, .
\label{eq:bc2}
\end{equation}

The particle propagation at all times and at each position in space depends on the diffusion coefficient $D(p,z,t)$, which is in turn affected by the turbulence excited by the particles. In the following we only consider resonant wave-excitation and scattering, namely a CR of momentum $p$ only excites waves of wavenumber $k=1/r_L(p)$ where $r_L(p)=p c /(e B_0)$ is the particle Larmor radius in the background field $B_0$ ($e$ is the proton charge). We further assume and {\it a posteriori} verify that the level of turbulence stays small ($\delta B\ll B_0$ at all times), and treat the problem within the framework of quasi-linear theory. (In fact waves may grow non-resonantly faster than the Alfv\'en waves considered here. In \S \ref{sec:conclude} we discuss the implications of such non-resonant growth for CR confinement in the near source region). Under these assumptions one can write~\citep{bell1978}:
\begin{equation}
D(p,z,t) = \frac{1}{3} r_L(p) \frac{v(p)}{\mathcal{F}(k, z, t) | _{k = 1/r_L(p)}} \, .
\label{eq:diffC}
\end{equation}
where $\mathcal{F}(k, z, t)$ is the turbulent magnetic energy density per unit logarithmic bandwidth of waves with wavenumber $k$, normalized to the background magnetic energy density $B_0^2/(8\pi)$, i.e. 
\begin{equation}
\frac{\delta B^2(z,t)}{8\pi} = \frac{B_0^2}{8\pi} \int \mathcal{F}(k, z, t) d\ln k \, .
\end{equation}
The evolution of the self-generated turbulence $\mathcal{F}(k, z, t)$ is described by the following wave equation
\begin{equation}
\frac{\partial \mathcal{F}}{\partial t} + u \frac{\partial \mathcal{F}}{\partial z} = (\Gamma_{CR} - \Gamma_D) \mathcal{F}(k, z, t) \, ,
\label{eq:calF}
\end{equation}
which we solve in the flux tube $0 \leq z \leq L_c$ where $u=v_A$. The evolution of $\cal F$ is determined by the competition between wave excitation by CRs at a rate $\Gamma_{CR}$ and wave damping, which occurs at a rate $\Gamma_D$. The non-linearity of the problem resides in the fact that the growth rate depends on the CR distribution function $f(p,z,t)$ as~\citep{skilling1971}
\begin{equation}
\Gamma_{CR} = \frac{16 \pi^2}{3} \frac{v_A}{\mathcal{F} B_0^2} \left |p^4 v(p) \frac{\partial f}{\partial z}\right|_{p = qB_0/(kc)} \, .
\label{eq:GammaCR}
\end{equation}
The value of the wave energy density $\cal F$ is regulated by the balance between growth and damping of waves: an increase in $\cal F$ leads to a decrease in the diffusion coefficient and to a longer confinement time of particles leaving the source.


The main damping mechanisms that can be at work in the situation we consider are: non linear damping~\citep{ptuskin2003}, ion-neutral damping~\citep{kulsrud1969,Zweibel1982} and damping due to pre-existing magnetic turbulence~\citep{farmer2004}. 

Non-linear Damping (NLD hereafter) is due to wave-wave interactions. Its rate can be written as~\citep{ptuskin2003}
\begin{equation}
\Gamma_{\rm NLD}(k,z,t) = (2 c_k)^{-3/2} k \, v_A \sqrt{\mathcal{F}(k,z,t)} \, ,
\label{eq:nld}
\end{equation}
where $c_k \approx 3.6$. We notice that this expression also well describes the evolution of a Dirac-delta function in $k$-space towards a Kolmogorov spectrum of waves. 

Ion neutral damping (IND hereafter) is a damping mechanism that only operates in partially ionized plasmas. The physical process causing wave energy dissipation is the viscosity produced by charge exchange interactions between ions and neutrals that cause former neutral particles to sudden participate in hydromagnetic phenomena when they become ionized. 
Hence, the IND  depends on how the wave's frequency, $\omega_k \equiv v_A k$, compares with the ion-neutral collisional frequency defined as $\nu_{in} \equiv n_n \langle \sigma v \rangle =  8.4 \times 10^{-9} (n_n/ {\rm cm^{-3}})(T/10^4 \rm K)^{0.4}\, s^{-1}$,  with $T$ the plasma temperature \citep{kulsrud1971}. We obtain the IND  rate numerically solving equation (A.4) from~\cite{Zweibel1982} \citep[see also][]{Nava2016} which also depends on the ratio between ion and neutral densities, $f_{in} = n_i / n_n$. When $f_{in} \gg 1$, a condition which applies to all cases analysed in this work,  such a solution is very well approximated by the following expression:
\begin{equation}  \label{eq:IN}
\Gamma_{\rm IN} \simeq 
	\frac{\nu_{in}}{2} \frac{\omega_k^2}{\omega_k^2 + f_{in}^2 \nu_{in}^2}  \,.
\end{equation}
We notice that in the large frequency limit, i.e. $\omega_k \gg \nu_{in} f_{in}$, ions and neutrals are not well coupled and the damping rate becomes independent of wave frequency, namely $\Gamma_{\rm IN} \simeq \nu_{in}/2$. On the contrary, in the low frequency limit, $\omega_k \ll \nu_{in} f_{in}$, we have $\Gamma_{\rm IN} \simeq \omega_k^2 /(2 f_{in}^2 \nu_{in}) \ll \nu_{in}/2$. In other words the damping is less effective because ions and neutrals are well coupled and oscillate together.

Finally, the last damping mechanism we consider has been proposed by~\cite{farmer2004}. These authors suggested that the waves generated by the CR streaming instability can be damped by interaction with pre-existing MHD turbulence. The damping rate associated with this process reads:
\begin{equation}
\Gamma_{\rm FG} = \frac{k v_A}{\sqrt{k L_{\rm MHD}}} 
\label{eq:FG}
\end{equation}
where $L_{\rm MHD}$ is the characteristic scale of the background MHD turbulence. 

Before presenting the results of our detailed calculations we think it useful to provide estimates of the relative importance of the various terms appearing in equation~(\ref{eq:calF}). As reference values for the parameters of the problem we assume $ B_0 = 3 \, \mathrm{\mu G}$ and $n_i = 0.45 \, \mathrm{cm^{-3}}$. Let us further focus on the wavelength corresponding to the case of 10 GeV particles, since these are the bulk of the CRs we will treat. At very early times (say, close to $t=0$), $\mathcal{F}(10 \mathrm{GeV}) \sim \mathcal{F}_g(10 \mathrm{GeV}) \sim 10^{-6}$, where $\mathcal{F}_g$ represents the Galactic turbulence. In addition, using the expression for $f(t=0)$ in equation~\eqref{eq:ic}, we can approximate $|\partial f/ \partial z|  \sim |2q_0/z_0|$. We then obtain, as an order of magnitude: 
\begin{equation}
\Gamma_{\rm CR}(10\ \mathrm{GeV}) \sim 2 \times 10^{-5} \mathrm{s^{-1}}\, .
\label{eq:GammaCR2}
\end{equation}
This value has to be compared with the damping rate associated to the different mechanisms listed above.

From equation~(\ref{eq:nld}), we have, in numbers,
\begin{align}
\Gamma_{\rm NLD}(E,z,t) =  4.7 &\times 10^{-9} \sqrt{\mathcal{F}(E,z,t)} \times \left ( \frac{B_0}{3 \mathrm{\mu G}} \right)^2   \nonumber \\
&\times \left( \frac{E}{10 \mathrm{GeV}} \right)^{-1} \left( \frac{n_i}{0.45 \mathrm{cm^{-3}}} \right)^{-1/2}  \mathrm{s^{-1}} \, ,
\end{align}
which implies that, close to $t=0$, when $\mathcal{F} \sim \mathcal{F}_g \sim 10^{-6}$,
\begin{equation}
\Gamma_{\rm NLD}(10\ \mathrm{GeV}) \sim 4.7 \times 10^{-12} \mathrm{s^{-1}} \ll \Gamma_{CR}(10\ \mathrm{GeV})\, .
\label{eq:nld2}
\end{equation}

In order to estimate the importance of the damping due to ion-neutral friction one has to specify the phase in which the ISM is in the region surrounding the SNR that is being described. The ISM exists in different phases, each with its own characteristics in terms of density and temperature. For our purposes we can focus on warm neutral medium (WNM), warm ionized medium (WIM) and hot ionized medium (HIM). For propagation of CRs on Galactic scales, it is likely that the HIM is the most important in that most of the propagation volume is in fact filled with such gas. On the other hand, in describing propagation of CRs in the near source regions, the situation may be quite different. Following \cite{ferriere2001} one can assume that the WIM is made of mostly ionized gas with density $\sim 0.45 \, \mathrm{ cm^{-3}}$ and a residual density of neutral gas with density $\sim 0.05 \, \mathrm{ cm^{-3}}$. As discussed by \cite{dangelo2016}, even this small fraction of neutral hydrogen would lead to severe IND in the near source regions. On the other hand,  \cite{ferriere1998} discussed the possibility that for gas temperature of $\sim 8000$ K a large fraction of this neutral gas may be made of helium rather than hydrogen. This latter picture would have prominent consequences in terms of IND because the cross section for charge exchange between H and He is about three orders of magnitude smaller\footnote{We use data retrieved by the Atomic and Molecular data service of the International Atomic Energy Agency, website: \texttt{https://www-amdis.iaea.org/ALADDIN/}.} than for neutral and ionized H, so that the corresponding damping rate would be greatly diminished. An upper limit to the density of neutral hydrogen can be written as $\lesssim 6\times 10^{-2} n_{i}$ \cite[]{ferriere1998}.

Using such limit in equation~(\ref{eq:IN}), one finds that, at the wavelengths that are relevant for 10 GeV particles, one has
\begin{equation}
\Gamma_{\rm IND}(10\ \mathrm{GeV}) \sim 4.5 \times 10^{-10} \; \mathrm{s^{-1}} \,,
\label{eq:in2}
\end{equation}
which may be overcome by the wave growth induced by the streaming of CRs.

It is worth mentioning that $H_{\alpha}$ observations of several type Ia SNRs often suggest neutral fractions that are close to $\sim 0.5$ \cite[see, e.g.,][]{2013SSRv..178..633G}, higher than those used above. However, one should keep in mind that 1) the region immediately surrounding the supernova may still have remnants of the cloud where the star formed from and 2) Balmer emission is preferentially observed from regions where the density of neutral hydrogen is high. Hence, it is understandable that the neutral fractions averaged over lines of sight \cite[]{ferriere2001} are typically much lower than suggested by observations of $H_{\alpha}$ emission from supernova shocks. We chose to adopt average values estimated in the ISM essentially because the scales that are relevant for this problem are much larger than the immediate surroundings of the SNR.

Finally, as far as the damping due to external turbulence is concerned, assuming $L_{\rm MHD}\sim L_{c}\sim 100$ pc, from equation~(\ref{eq:FG}) we estimate:
\begin{equation}
\Gamma_{\rm FG}(10\ {\rm GeV})= 1.2\times 10^{-11}  \mathrm{s^{-1}}  \,,
\label{eq:FG2}
\end{equation}
which is again negligible with respect to the growth rate at early times as estimated in equation~\ref{eq:GammaCR2}.
The obvious conclusion is that, at least at the beginning of the escape history of CRs into the ISM, the instability is expected to efficiently grow.

In order to study what happens at later times, the full time-dependent system of equations described above must be solved. The numerical procedure adopted to solve equation~\ref{eq:transp} and equation~\ref{eq:calF} is based on a finite difference method for the discretization of partial derivatives (both in space and time) and a backward integration in time for equation~\eqref{eq:transp}. 

Given that neutral friction is likely the most effective wave damping mechanism, depending on the ionization fraction of the medium surrounding the source, we have tried to assess its importance considering four different types of ISM \cite[in analogy with what was done by][]{dangelo2016}: (1) fully ionized medium with ion density $n_i=0.45~\rm cm^{-3}$; (2) partially ionized medium with ion density $n_{i}=0.45~\rm cm^{-3}$ and neutral density $n_{n}=0.05~\rm cm^{-3}$; (3) partially ionized medium with $n_i=0.45~\rm cm^{-3}$ and $n_n=0.03~\rm cm^{-3}$; (4) rarefied totally ionized medium with $n_{i}=0.01~\rm cm^{-3}$.

\begin{figure}
\centering
\includegraphics[width=1.0\columnwidth]{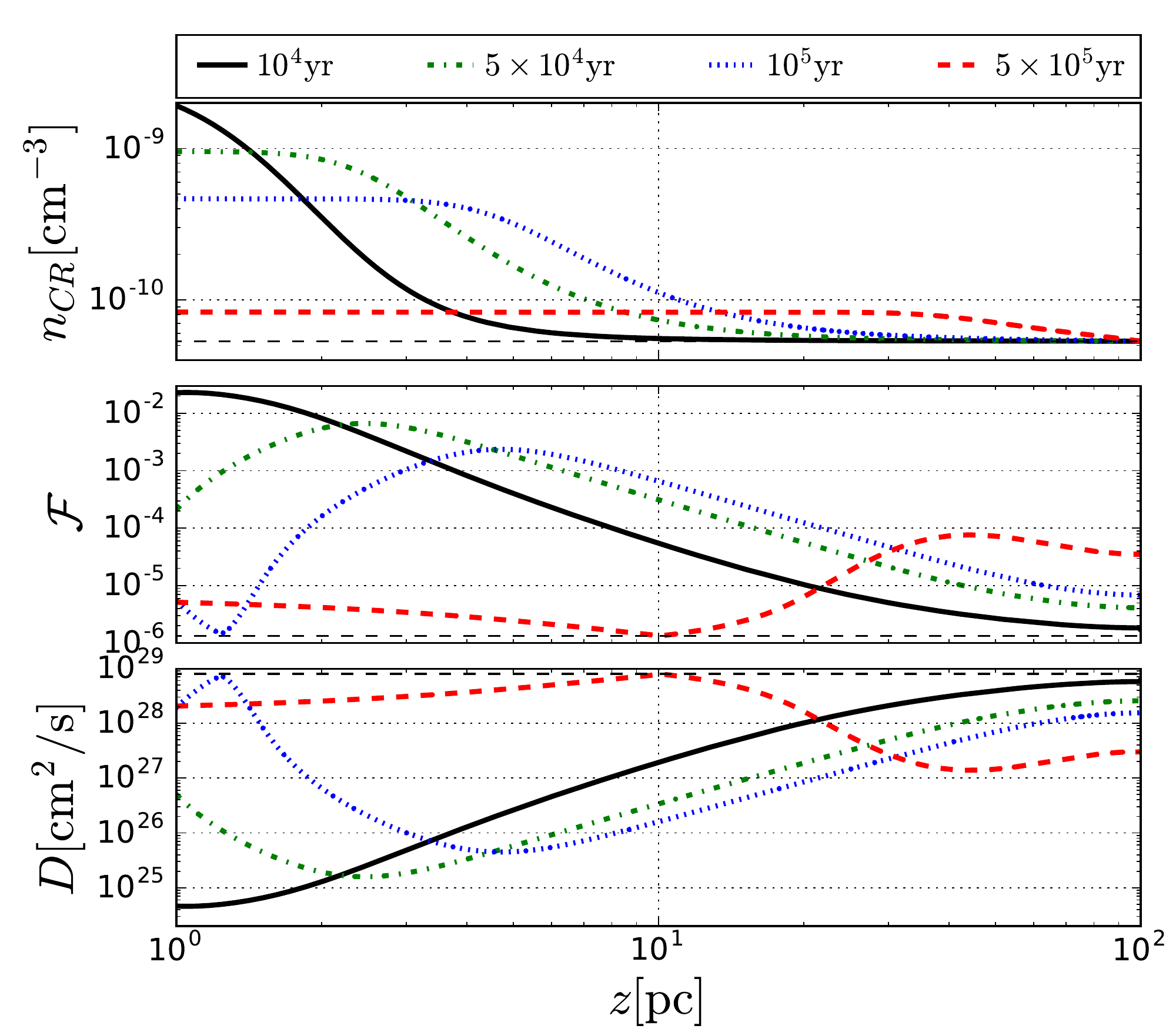}
\caption{Spatial profile of the CR density, turbulence energy density and diffusion coefficient at an energy of 10 GeV and at four different times after the particle release by the source, as specified on top of the figure. \emph{Top panel}: CR density $n_{CR}= (4/3)\pi p^3 f$, the black dashed line represents the Galactic distribution $n_g = (4/3)\pi p^3 f_g$, with $f_g(p)$ defined in equation~\eqref{eq:fg}. \emph{Center panel}: Self-generated turbulence $\mathcal{F}$, the black dashed line represents the Galactic turbulence $\mathcal{F}_g$. \emph{Bottom panel}: CR diffusion coefficient $D$, the black dashed line represents the Galactic diffusion coefficient $D_g$ \protect\cite[as calculated by][]{ptuskin2009}. Here we assumed a fully ionized medium with ion density $n_i = 0.45$ cm$^{-3}$.}
\label{fig:CR_dist}
\end{figure}

In Fig.~\ref{fig:CR_dist} we show the spatial profiles of the relevant quantities describing the propagation of 10 GeV particles, at four different times after release of the freshly accelerated particles in the ISM. The curves have been obtained as numerical solutions of equation~(\ref{eq:transp}), (\ref{eq:diffC}) and (\ref{eq:calF}) in the case of a fully ionized medium with density $n_i = 0.45 \, \mathrm{cm^{-3}}$. 
The first thing to notice in Fig.~\ref{fig:CR_dist} is that the density of 10 GeV particles, $n_{CR} = (4/3)\pi p^3 f$, exceeds the average Galactic density $n_g = (4/3)\pi p^3 f_g$  by more than one order of magnitude at least for $t\lesssim 10^5 \, \mathrm{yr}$ in a region $\sim 5 \, \mathrm{pc}$ around the parent source. Even after $5 \times 10^5 \, \mathrm{yr}$ the particle density is still larger than the Galactic density for $z$ up to 30 pc. This time is much longer than the standard diffusion time $\tau_d = L_c^2/D_g$ at $E = 10$ GeV which is of the order $\approx 4 \times 10^4 \, \mathrm{yr}$.

The maximum level of amplification reached by the magnetic turbulence is $\mathcal{F}\lesssim 10^{-2}$ (as can be seen from the central panel in Fig.~\ref{fig:CR_dist}). This is about three orders of magnitude higher than the background turbulence $\mathcal{F}_g$, but still below the limit of validity of quasi-linear theory, which requires $\mathcal{F} \ll 1$. When the bulk of particles escapes from the region, the level of turbulence drops to $\mathcal{F}_g$. The behavior of $\mathcal{F}$ at small $z$ is due to the advection term $v_A \partial f/\partial z$ and the associated boundary condition~\eqref{eq:bc1}: advection of the particles with self-generated waves cause a depletion of particles in the inner region ($z\to 0$), which leads to a small gradient in the density of CRs in the same region. In turn this gradient causes the growth of Alfv\'en waves moving towards $z\sim 0$. This is the reason why the spatial profiles of $\mathcal{F}$ shows the dips and peaks that are visible in Fig. ~\ref{fig:CR_dist}. A note of caution is required here: since the waves excited by the advection-induced gradient move toward the origin, it is likely that in this region the net Aflv\'en velocity gets lower or even vanishes, depending on the compensation due to waves traveling in the opposite direction. This effect is hard to take into account in a quantitative manner. In any case, since this phenomenon is limited to the region very close to the source, its effect on the confinement time and grammage accumulated by particles is negligible, at least for WIM gas density of order $\sim 1 ~ \rm cm^{-3}$. For a rarefied medium, the Alfv\'en speed is larger and the effect of advection becomes more important. On the other this is also the case when no appreciable effect on the grammage is expected (see discussion below).
In the bottom panel of Fig.~\ref{fig:CR_dist} we report the spatial diffusion coefficient $D$, which clearly shows a profile that reflects the dips and peaks of  $\mathcal{F}$, being $D\propto {\cal F}^{-1}$: at the location where the level of amplification is maximum, $D$ reaches its minimum of $\approx 4 \times 10^{25} \, \mathrm{cm}^2/\mathrm{s}$, more than three orders of magnitude lower than the Galactic diffusion coefficient (dashed black line).   

We further proceed to discuss the confinement time of particles around their sources. Due to the non-linear processes involved, this calculation is not straightforward. In the case where a burst of particles is injected a $t=0$ with a preassigned diffusion coefficient equal to $D_g(p)$, we find that about $89\%$ of the total injected particles $N_{CR}^{inj}(p) = 2\pi L_c R_{SN}^2q_0(p)$ leave the box within the classical diffusion time $\tau_d=L_c^2/D_g$. We used this result to estimate the escape time in the nonlinear case. For a given particle momentum, the number of freshly accelerated particles inside the tube, $N_{CR}(p,t)$, is calculated from the total amount of particles in the tube $2\pi R_{SN}^2\int_{0}^{L_c} dz f(p,z,t)$ after subtraction of the background term $2\pi R_{SN}^2L_cf_g(p)$. The escape time $t_{\rm esc}(p)$ is defined as the time at which $N_{CR}(p, t_{\rm esc}(p)) = 0.11 N_{CR}^{inj}(p)$. 
We should bear in mind that we assumed the duration of the escape time of accelerated particles from the source to be much shorter than the propagation time in the near source region. For typical values of the parameters this assumptions remains satisfied for energies $\lesssim 10$ TeV, if one assumes a release time for an average SNR of about $10^{4}$ years \cite[see, e.g.][]{Caprioli+2009}.

\begin{figure}
\centering
\includegraphics[width=1.0\columnwidth]{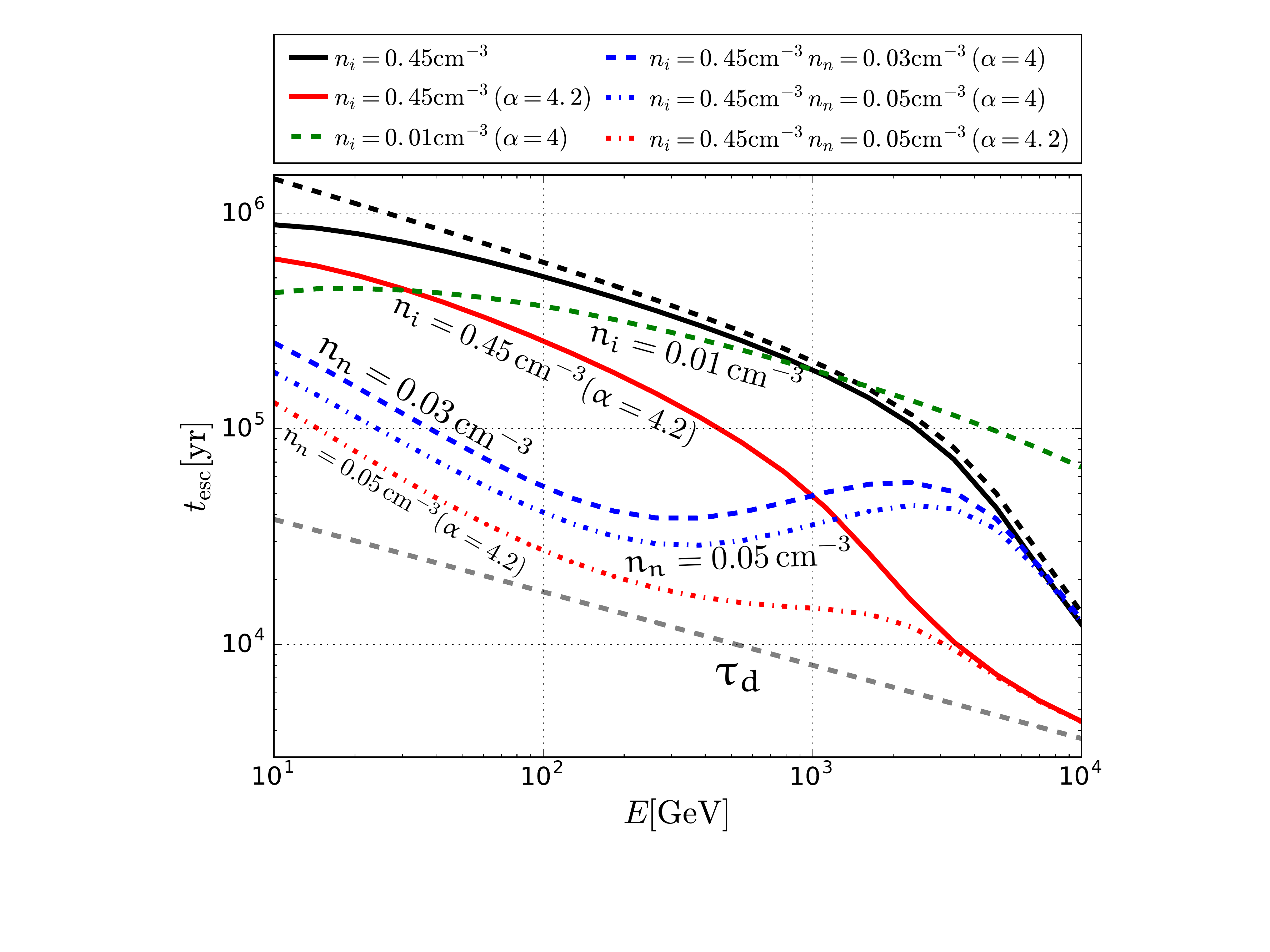}
\caption{CR escape time $t_{\rm esc}$ (see text for exact definition) as a function of particle energy for four different types of ISM: (1) no neutrals and $n_i = 0.45\,  \mathrm{cm^{-3}}$ (black solid line); (2)$n_{n}=0.03 \, \mathrm{cm^{-3}}$ and $n_i = 0.45 \, \mathrm{cm^{-3}}$ (blue dashed line); (3) $n_n = 0.05 \, \mathrm{cm^{-3}}$ and $n_i = 0.45 \, \mathrm{cm^{-3}}$ (blue dot-dashed line); (4) no neutrals and $n_i = 0.01 \, \mathrm{cm^{-3}}$ (green dashed line). 
The black dashed line represents case (1) but without the advection term in both equations~\eqref{eq:transp} and \eqref{eq:calF}.
The solid red line represents case (1), but with a steeper injection spectrum, $\alpha = 4.2$; analogously, the dot-dashed red line corresponds to case (3) with $\alpha=4.2$. For all other curves the slope is $\alpha = 4$. The grey dashed line represents the standard diffusion time $\tau_d = L_c^2/D_g = 8.3 \times 10^4 (E_{\mathrm{GeV}})^{-1/3} \, \mathrm{yr}$.}
\label{fig:tesc}
\end{figure}

We plot the escape time as a function of particle energy in Fig.~\ref{fig:tesc}, considering the four different types of ISM listed above. For case (1) and case (3) also a steeper power-law spectrum at injection has been considered, with $\alpha=4.2$. The  escape time in the case of propagation in the Galactic diffusion coefficient, $\tau_d = 8.3 \times 10^4 (E_{\mathrm{GeV}})^{-1/3} \, \mathrm{yr}$, is also plotted for comparison, as a grey dashed line.
In case (1), where neutrals are absent (solid black line), the escape time is about two orders of magnitude longer than the diffusion time $\tau_d$ for energies up to few $\mathrm{TeV}$; for higher energies, the efficiency of streaming instability decreases, and $t_{\rm esc}$ starts approaching $\tau_d$, although, even at 10 TeV we find $t_{\rm esc}(10 \, \mathrm{TeV}) \sim 6 \times \tau_d(10 \, \mathrm{TeV})$. 
In case (4), where the medium is fully ionized but more rarefied ($n_i = 0.01 \, \mathrm{cm^{-3}}$, dashed green line), the escape time is roughly the same as the previous case for energies in the range $10^2 - 10^3 \, \mathrm{GeV}$. At lower energies, some difference arises due to the relative importance acquired by advection: in the lower density medium, the Alfv\'en speed is larger and advection makes escape of low energy particles faster.
This finding requires some additional comments: as we discussed above, the self-generated waves move in the direction of the CR gradient. When advection is included in the calculation, the CR density at given energy develops a positive gradient when the source is no longer injecting CRs, and as a result the self-generated waves tend to move in the direction of the source. This situation would lead to waves moving in both direction that in turn would reduce the effective wave speed and possibly make it vanish. In other words, when the source is switched off, it is likely that the effective wave speed to be used in the transport equation is much smaller than the Alfv\'en speed and perhaps it equals zero. Since this effect cannot be introduced in the calculations self-consistently, in Fig. \ref{fig:tesc} we also show the confinement time resulting for case (1) but setting the advection speed to zero (black-dashed line). As expected, at low energies the decrease in the confinement time disappears. The effect is more prominent in the case of low density fully ionized background plasma (not shown in the figure). 

The importance of self-confinement is effectively reduced when neutrals are present in the medium \cite[see][]{dangelo2016,Nava2016}. The dashed black line and the dot-dashed black line represent the escape time for case (2) with $n_n = 0.03 \, \mathrm{cm^{-3}}$ and (3) with $n_n = 0.05 \, \mathrm{cm^{-3}}$, respectively. Ion-Neutral damping decreases the level of turbulence especially in the energy range $[0.1 - 1] \, \mathrm{TeV}$. However, the escape time remains longer than the diffusion time in the galactic diffusion coefficient for all energies, and in particular for $E \le 50 \, \mathrm{GeV}$ and $E \sim \mathrm{few} \, \mathrm{TeV}$. This is because the damping rate of the IND remains roughly constant up to $E \approx 500 \, \mathrm{GeV}$ and then decreases $\propto E^{-1}$, so that the quenching of growth is maximum in the energy range $[10^2 - 10^3] \, \mathrm{GeV}$. 
We have also analysed the effect of a steeper CR injection spectrum; the red lines, solid and the dot-dashed, correspond to the injection of a particle spectrum with $\alpha = 4.2$ in case (1) and case (3), respectively. In both these situations, the escape time is close, at low energies, to that found for $\alpha=4$. However, at high energies, the growth of streaming instability is considerably reduced and $t_{\rm esc}$ becomes close to $\tau_d$.
 
In the next section we discuss some possible observational signatures of effective self-confinement of CRs in the vicinity of their sources. 

\section{Gamma-ray emission}
\label{sec:gamma}

\subsection{Emission from an individual near-source region}
\label{sec:gamma1}
As explained in Sec.~\ref{sec:Intro}, CRs can interact with the ambient gas in the near source region via inelastic scattering and produce $\pi^0$-mesons which, in turn, decay into $\gamma$-ray photons in the energy range $[1 - 10^3] \, \mathrm{GeV}$, i.e. $p_{CR} + p_{ISM} \to \pi^0 \to 2 \gamma$ \cite[see][and references therein]{aharonian2000}. 

The self-confinement process can give rise to the formation of a halo of gamma radiation. According to the description adopted in the previous section, we expect this halo to be elongated in the direction of the ambient magnetic field. In the following we estimate the contribution of these extended halos to the diffuse galactic $\gamma$-ray background.
In order to do this, we start by evaluating the emissivity associated to a single SNR halo, then assume some distribution of SNRs in the Galaxy, and finally proceed to summing up all the contributions from SNR halos within given regions of space for direct comparison with available data.
Using the numerical solutions of the coupled equations~\eqref{eq:transp} and~\eqref{eq:calF} (as explained in Sec.~\ref{sec:CR_dist}), we calculate the time-dependent, spatially integrated, CR spectrum $J_{CR}(E,t) \equiv dN_{CR}/dE$ associated with a single SNR. Spatial integration is performed over a cylindrical flux tube of radius $R_{SN}$ and length $2 L_c$: 
\begin{equation}
J_p(E, t) = 2 \pi R_{SN}^2 \int_0^{L_c} dz \, f(E,z,t) \, ,
\label{eq:gemiss}
\end{equation}
where we have used the relation $4 \pi p^2 f(p) dp = f(E) dE$ and the symmetry of the problem around $z=0$. The volume-integrated gamma-ray emission $q_{\gamma}(E_{\gamma},t)$ is then computed using the $\delta$-function approximation detailed in the works by \cite{aharonian2000} and \cite{kelner2006}. 

\begin{figure}
\centering
\includegraphics[width=1.0\columnwidth]{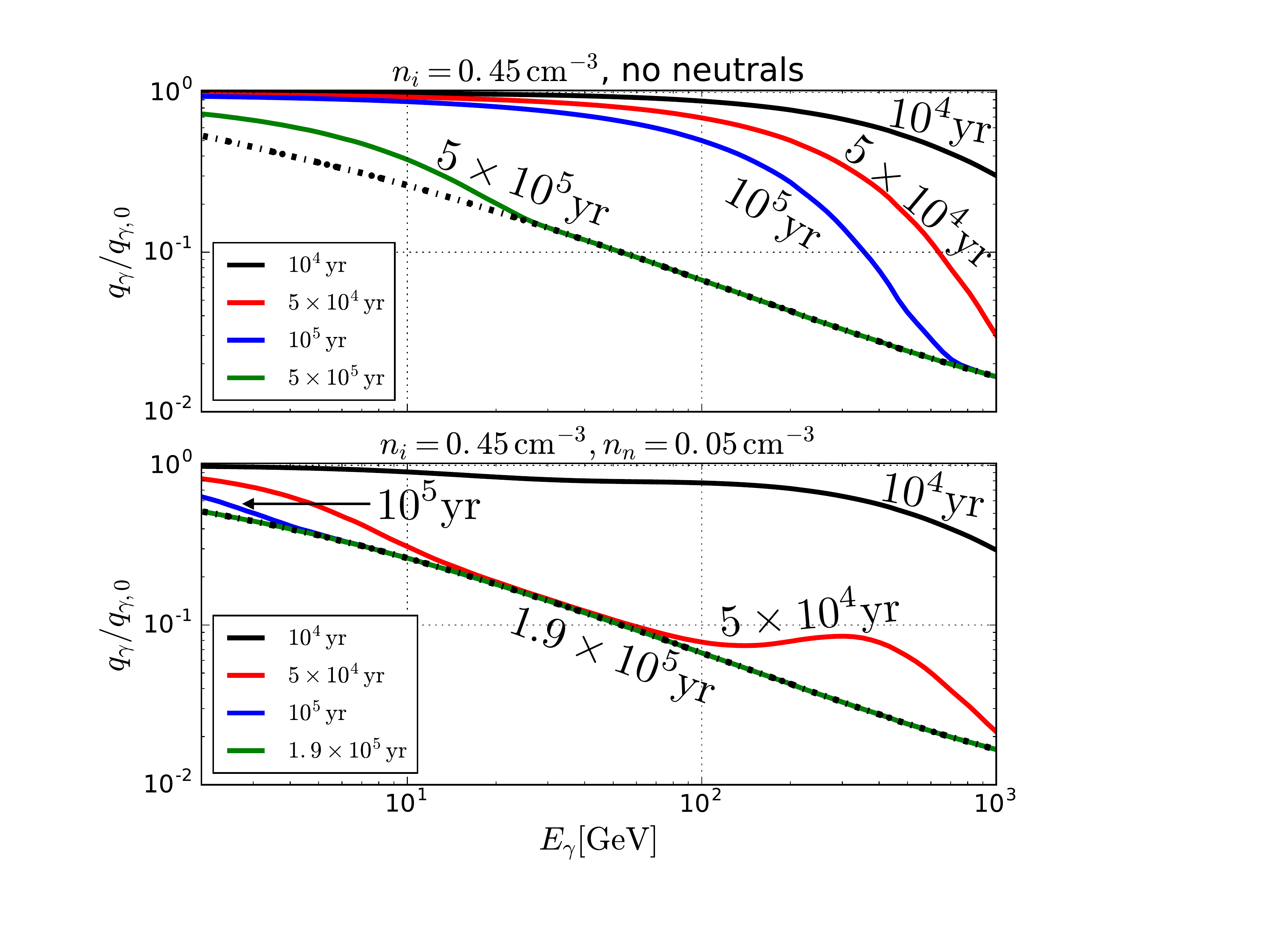}
\caption{Normalized $\gamma$-ray emission $q_{\gamma}(E_{\gamma},t)/q_{\gamma}(E_{\gamma},t = 0)$ (integrated over the flux tube volume) at four different times and for two different types of ISM, as written in the labels.
\emph{Top panel}: fully ionized medium with $n_i = 0.45 \, \mathrm{cm^{-3}}$.
\emph{Bottom panel}: partially ionized medium with $n_i = 0.45 \, \mathrm{cm^{-3}}$ and $n_n = 0.05 \, \mathrm{cm^{-3}}$. 
The black dot-dashed line represents the gamma-ray emission of the galactic CR pool integrated over the same volume ($f = f_g$ in equation~(\ref{eq:gemiss})).}
\label{fig:gamma_emissivity}
\end{figure}

In Fig.~\ref{fig:gamma_emissivity} we show the volume-integrated gamma-ray emission, $q_{\gamma}(E_{\gamma},t)$, at four different times during the SNR evolution, between $10^4$ and $few \times 10^5$ yr. The two panels refer to different ambient conditions, described as case (1) and case (3) in the previous section. All curves are normalized to their value at time $t=0$, with the notation $q_{\gamma}(E_{\gamma},t = 0) \equiv q_{\gamma,0}$. 
In the top panel, we present the results for fully ionized ISM with $n_i = 0.45 \, \mathrm{cm^{-3}}$ (case (1)). At time $t = 10^4 \, \mathrm{yr}$ CRs with $E \lesssim 10^3$ GeV are all confined within the box, which implies that for $E_{\gamma} \lesssim 10^2 \, \mathrm{GeV}$ one has $q_{\gamma} \simeq q_{\gamma,0}$. For particles below $10^3$ GeV, confinement is effective until around $10^5$ yr, so that the emissivity of photons with energy below 100 GeV, is almost constant until that time. On the other hand, at $t\approx 5 \times 10^5$ yr, even 10 GeV particles start to leave the near-source region, so that the local CR distribution function starts to approach its diffuse galactic counterpart at all relevant energies, and so does the gamma-ray emission. This is clearly seen from comparison of the solid green curve with the black dot-dashed curve in the top panel of Fig.~\ref{fig:gamma_emissivity}.
We notice that while we do not plot it explicitly, the emission of case (4), corresponding to a more rarefied but still fully ionized medium, shows the same time evolution.

The bottom panel of Fig.~\ref{fig:gamma_emissivity} shows the gamma-ray emission for case (3), i.e. for a partially ionized medium with $n_n = 0.05 \, \mathrm{cm^{-3}}$. When neutrals are present in the ambient gas, as shown in Fig.~\ref{fig:tesc}, the escape time is largely reduced with respect to the above case. As a consequence, the CR distribution function in the near source region approaches $f_g$ much more quickly after $10^4$ yr at all relevant energies. One interesting thing to notice is the difference in shape between the red curves in the two panels: in the case of a partially ionized medium, a bump arises in the gamma-ray emissivity between $10^2$ and $10^3$ GeV. This will give rise to a different integrated spectrum, as we discuss in Sec.~\ref{sec:results}.

\subsection{Total gamma-ray emission from the Galactic disc}
\label{sec:tot_flux}
In order to estimate the contribution of self-confined CRs to the diffuse gamma-ray emission, the second step to take is the sampling of the SNR distribution in the Galaxy, more precisely in the galactic disk. The Galactic disk is characterized by a diameter of $ \sim 30 \, \mathrm{kpc}$ and a thickness of $\sim 300 \, \mathrm{pc}$. Given the large difference between these two sizes, a reasonable approximation is that of an infinitely thin disk, which allows to treat the system as 2D. In the following we assume $R_{\rm disc} = 16.5$ kpc and use polar coordinates $R\in [0, R_{\rm disc}]$ and $\phi\in [0, 2\pi]$ to identify locations in the disk.
The SNR distribution is taken as uniform in the azimuthal coordinate, $\phi$, and according to that reported by \cite{green2015} in terms of galactocentric distance:
\begin{equation}
g_{SN,R}(R) = A \left( \frac{R}{R_{\odot}}\right)^{\alpha} \exp \left( -\beta \frac{R - R_{\odot}}{R_{\odot}}\right)\ ,
\label{eq:rad_dist}
\end{equation}
where $A = 1/\int_0^{R_{\rm disc}} g_{SN}(R) dR$ is a normalization constant, $\alpha = 1.09$, $\beta = 3.87$ and $R_{\odot} = 8.5 \mathrm{kpc}$.

An advantage of this particular distribution with respect to others available in the literature is that it gives a peak of the SNR density at $R \sim 5 \, \mathrm{kpc}$. Such an over-density of SNRs has been independently suggested by studies of the diffuse gamma-ray emission \citep{yang2016,fermi2016} also in the framework of non-linear diffusion models \citep{Recchia2016b}.
We have checked that the degeneracy in the parameters $\alpha$ and $\beta$, which are obtained by fitting the observed SNRs with a power-law/exponential model, does not have a significant impact on our results.

On average, a SN is expected to explode in our Galaxy every $30 \, \mathrm{yr}$. We assume that the time at which a SN explodes follows a Poisson distribution where the mean rate of explosion is $\mathcal{R}_{SN} = 1/30 \,  \mathrm{yr}^{-1}$. The sources that contribute to the total $\gamma$-ray emission are those with an age that does not exceed the maximum confinement time ($t_{\max}$), which we take equal to the confinement time of 10 GeV particles in each of the considered cases.

In the following we will compare our results with those of two recent studies of Galactic diffuse emission. The first work is that of \cite{yang2016}, where the authors consider 7 years of observations with {\it Fermi}-LAT and provide gamma-ray spectra integrated over small angular sectors around 3 different directions: the direction of the Galactic centre, the opposite one, and the orthogonal one.

In order to compare our results with those data, we had to switch from galactocentric coordinates to the galactic system of coordinates, which is centered on the Sun. In this reference frame, each source is identified by the coordinates $d$ and $l$, representing the distance from the Sun and the galactic longitude respectively. The latter is the angular distance from a line connecting the Sun with the Galactic centre. Once $d$ and $l$ have been introduced, the gamma-ray flux to compare with data is computed as:
\begin{equation}
\phi_{\Delta l}(E_{\gamma}, \Delta l) = \sum_{t_{\rm age} \leq t_{\max}} \sum_{l_1 \leq l \leq l_2} \frac{q_{\gamma}(t_{\rm age}, E_{\gamma})}{4 \pi d^2} \, ,
\label{eq:felix}
\end{equation}
where $\Delta l=l_1-l_2$ and $l_1$ and $l_2$ are the galactic longitudes describing a particular angular sector. The sum is calculated on all SNe exploded in the relevant field of view with the appropriate $t_{age}$, the time passed since explosion, which in turn defines the extent of the diffusion region around the remnant. It is worth noticing that, if the spatial density of SNe in the Galactic disc were constant, the sum in equation (\ref{eq:felix}) would be dominated (through only logarithmically) by nearby sources. A more realistic distribution of sources weakens this conclusion, although it remains true that local sources induce sizable fluctuations in the diffuse gamma ray background.

We also compare our results with those of another article, by the Fermi collaboration \citep{fermi2016}, where the spectral energy distribution at various Galactocentric distances is derived. More precisely, what the authors plot is the spectral energy distribution per $H$ atom, a quantity that depends on the particular model assumed for the ISM in which CRs propagate. This fact makes the comparison of our results with those data more subtle.

The relevant quantity for comparison is, in this case, the integral of the gamma-ray emissivity per unit hydrogen atom over an annular sector centered on the Galactic centre. Going back to galactocentric coordinates, we compute:
\begin{equation}
\phi_{\Delta R}(E_{\gamma}, \Delta R) = \sum_{t_{\rm age} \leq t_{\max}} \sum_{R_1 \leq R \leq R_2} 
			\frac{q_{\gamma}(t_{\rm age}, E_{\gamma})}{4 \pi n_i \Delta V} \, ,
\label{eq:fermi}
\end{equation}
where $\Delta R = R_2 - R_1$ and $\Delta V = \pi h_d (R_2^2 - R_1^2)$, with $h_d = 300 \, \mathrm{pc}$ the height of the Galactic disc.

The quantity in equation~\eqref{eq:fermi} holds different information with respect to the one in equation~\eqref{eq:felix}, in that now the contribution of the different sources is not weighted with their distance from the Sun. 

\section{Comparison with the Galactic diffuse emission}
\label{sec:results}
We are now ready to compare our results with those derived from observations by \cite{yang2016} (see their figure 1) and \cite{fermi2016} (see their figure 7).

In order to take into account the statistical fluctuations of the spatial and temporal SNR distribution we considered 100 different realizations of such distribution in the Galaxy. For each realization we computed $\phi_{\Delta l}$ and $\phi_{\Delta R}$ according to equations~\eqref{eq:felix} and \eqref{eq:fermi} and then we averaged over the different realizations. In the following figures we show  our results in the form of a band centred on the average value we found and with a width representative of the statistical uncertainty: the band width is such that 68\% of the considered realizations result in gamma ray fluxes within such band.

\begin{figure*}
\centering
\includegraphics[width=1.0\columnwidth]{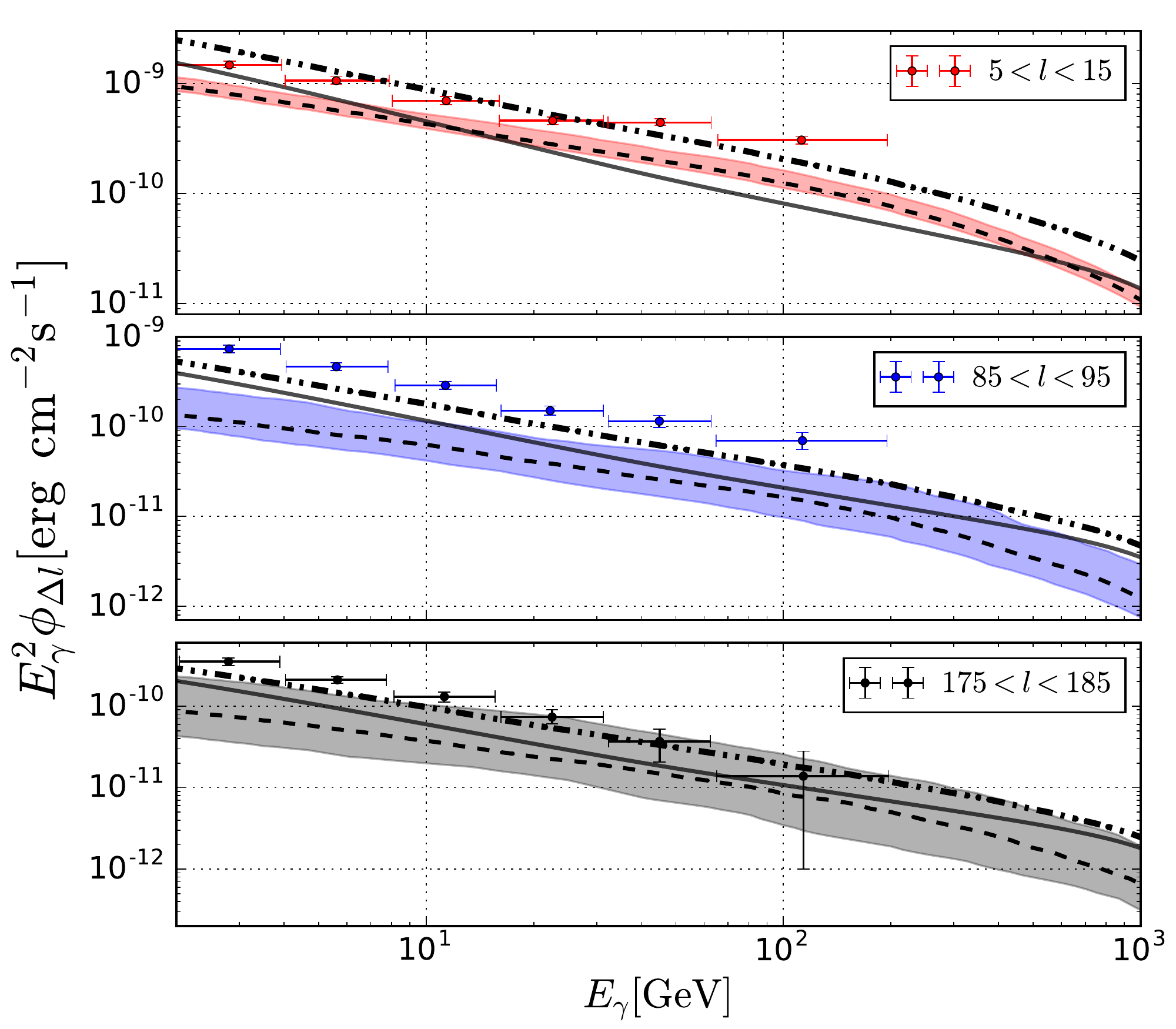} 
\includegraphics[width=1.0\columnwidth]{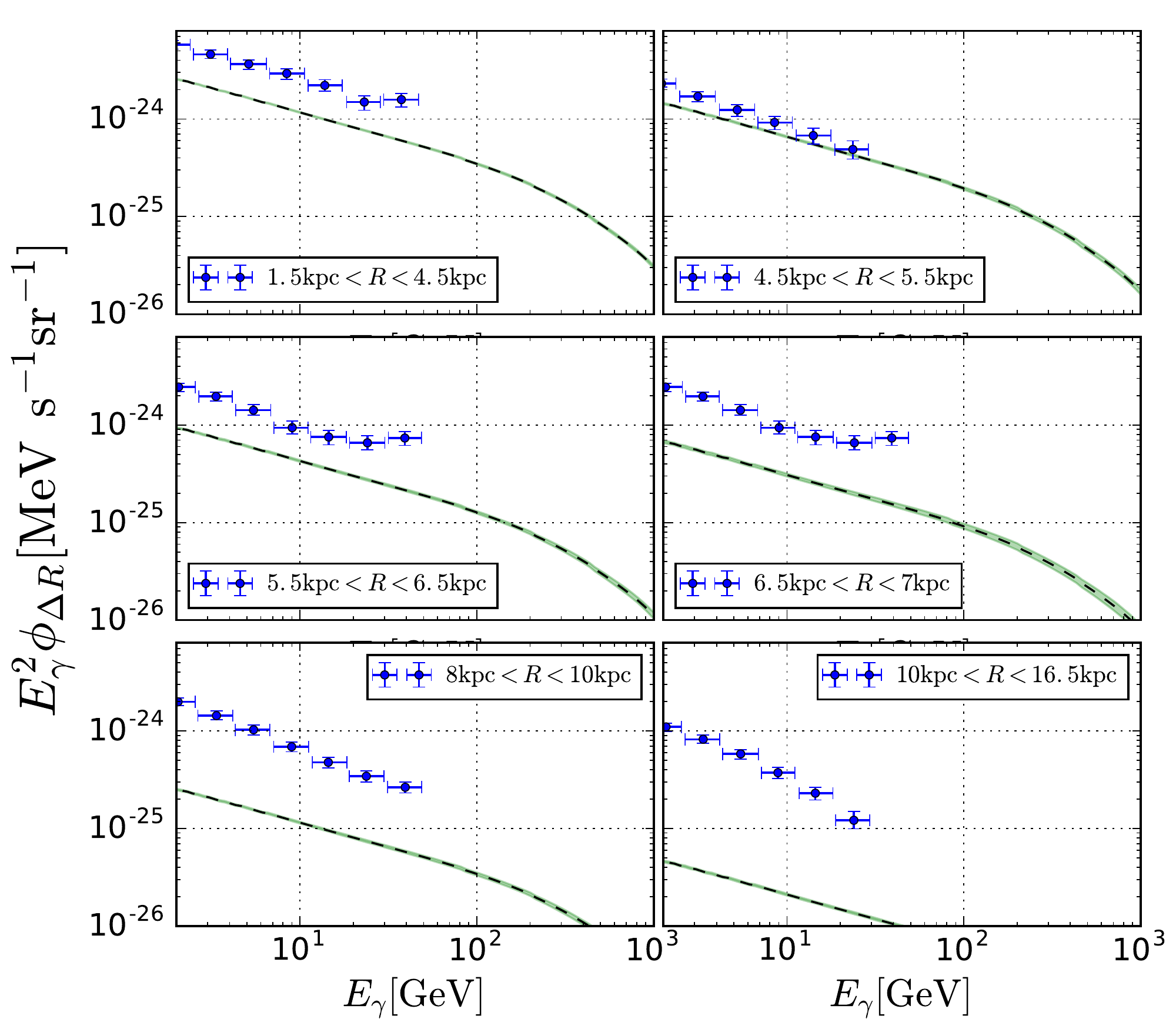} 
\caption{Case (1): fully ionized medium with $n_i = 0.45 \,  \mathrm{cm}^{-3}$. \emph{Left panel}: spectral energy distribution of the total gamma-ray emission $E_{\gamma}^2 \times \phi_{\Delta l}$ in three different angular sectors, $5^{\circ}\le l \le 15^{\circ}$ (top), $85^{\circ}\le l \le 95^{\circ}$ (centre) and $175^{\circ}\le l \le 185^{\circ}$ (bottom). The black dashed line corresponds to the mean value, while the band corresponds to a confidence level of $68 \%$. The solid line corresponds to the mean diffuse gamma-ray emission, while  the dash-dot-dot line corresponds to the sum of the average gamma-ray emission produced by self-confined CRs with the mean diffuse gamma-ray emission (sum of the dashed and the solid lines).
The data points are from \protect\cite{yang2016}. \emph{Right panel}: the spectral energy distribution of the total gamma-ray emission per $H$ atom, $E_{\gamma}^2 \times \phi_{\Delta R}$, in six different ring sectors (see labels). The black dashed line corresponds to the mean value, while the band corresponds to a confidence level of $68 \%$. The data points are from \protect\cite{fermi2016}.
}
\label{fig:ni045nn0}
\end{figure*}

In Fig.~\ref{fig:ni045nn0} we compare the analyzed data with our results in case of a fully ionized medium with $n_i = 0.45 \, \mathrm{cm^{-3}}$.
In the left panel we show the spectral energy distribution of the total $\gamma$-ray emission in the three different angular sectors considered by \cite{yang2016}. The top left panel is for the sector $\Delta l = 5^{\circ}\le l \le 15^{\circ}$, which is in the direction of the central region of the Galaxy, though avoiding the Galactic centre. In this sector, the number of SNRs that contribe to the gamma-ray emission averages to $\sim 700$. The central left panel is for the sector $\Delta l = 85^{\circ}\le l \le 95^{\circ}$, namely a line of sight roughly perpendicular to the Galactic centre. In this case, the contributing SNRs are $\sim 350$ on average, about half with respect to the previous case. This explains the increased width of the uncertainty band: the statistical error is larger with respect to the previous case due to the reduced number of contributing sources. In the sector $\Delta l = 175^{\circ}\le l \le 185^{\circ}$ (bottom left panel), corresponding to a line of sight towards the outskirts of the Galaxy, the uncertainty is even more severe, due to a mean number of sources of only $\sim 15$.

In all cases, our curves have a slope similar to the one that can be inferred from the data (especially at low energies) and their normalization is not too far below the data. This latter fact suggests that CR self-confinement can provide a non-negligible contribution to the diffuse emissivity, especially in the direction of the galactic centre, where the majority of sources is located. In this sector, at $E_{\gamma} = 2 \, \mathrm{GeV}$ we find that the average value of the flux produced by trapped CRs is about $60\%$ of the observed flux. However this value, as well as the fact that in the bottom panel the observed flux is almost saturated by our prediction, should be interpreted with caution. These plots refer to a fully ionized medium, which is certainly not a realistic assumption if extended to the entire SNR population. Therefore, the results in this figure should be interpreted as upper limits to the contribution of self-confined particles to the diffuse gamma-ray emission. On the other hand, one should also notice that the density of gas in the central regions of the Galaxy might be somewhat larger and thereby increase the contribution of such regions to the diffuse background. Moreover, while in this article we focus on the role of resonant streaming instability, the non-resonant version of this instability \cite[]{2004MNRAS.353..550B} might play an important role in that it may grow faster at least during some stages of the CR propagation in the near source region. In such phases it might be easier to overcome damping and confine CRs more effectively. However it is very difficult to reliably estimate these effects (see discussion in \S\ref{sec:conclude}).

Our predictions are also compared with the diffuse gamma-ray background produced by the interaction of the CR pool with the ISM gas, shown in the left panel of Fig.~\ref{fig:ni045nn0} as the solid line. To perform this calculation we used the Galactic CR (proton) spectrum measured by AMS-02 and given by equation~\eqref{eq:fg} (assumed to be spatially constant in the whole Galaxy), while for the gas distribution of the ISM we have used the cylindrically symmetric model reported by \cite{moskalenkoAPJ2002} (see their appendix A) for both $\rm H_I$ and $\rm H_{II}$. In Fig.~\ref{fig:ni045nn0} we also report as an orange dot-dashed line the sum of the diffuse gamma-ray emission plus the mean contribution from trapped CRs (dashed curves). It is worth noticing that the contribution due to the mean CR distribution alone cannot adequately account for the data (both in normalization and slope), especially in the sector close to the Galactic centre as already reported by \cite{fermi2016} and \cite{yang2016}.

In the right panel of Fig.~\ref{fig:ni045nn0} we show the spectral energy distribution of the total $\gamma$-ray emission per $H$ atom in six different annular regions. In each sector the mean number of SNRs is rather large, around $[10^3 - 2\times10^3]$, which explains why the width of the band appears so small when compared with the plots on the left. In terms of comparison between model results and data, the first thing one notices is that again self-confinement can in principle provide a non-negligible contribution to the diffuse gamma-ray emission, especially in the inner part of the Galaxy, where SNRs are more abundant. Once again one must not worry about the fact that in the sector $4.5{\rm kpc}<R<5.5 {\rm kpc}$ our emissivity almost saturates the data, because once again this situation corresponds to an upper limit to what the contribution of this process can be.

\begin{figure*}
\centering
\includegraphics[width=1.0\columnwidth]{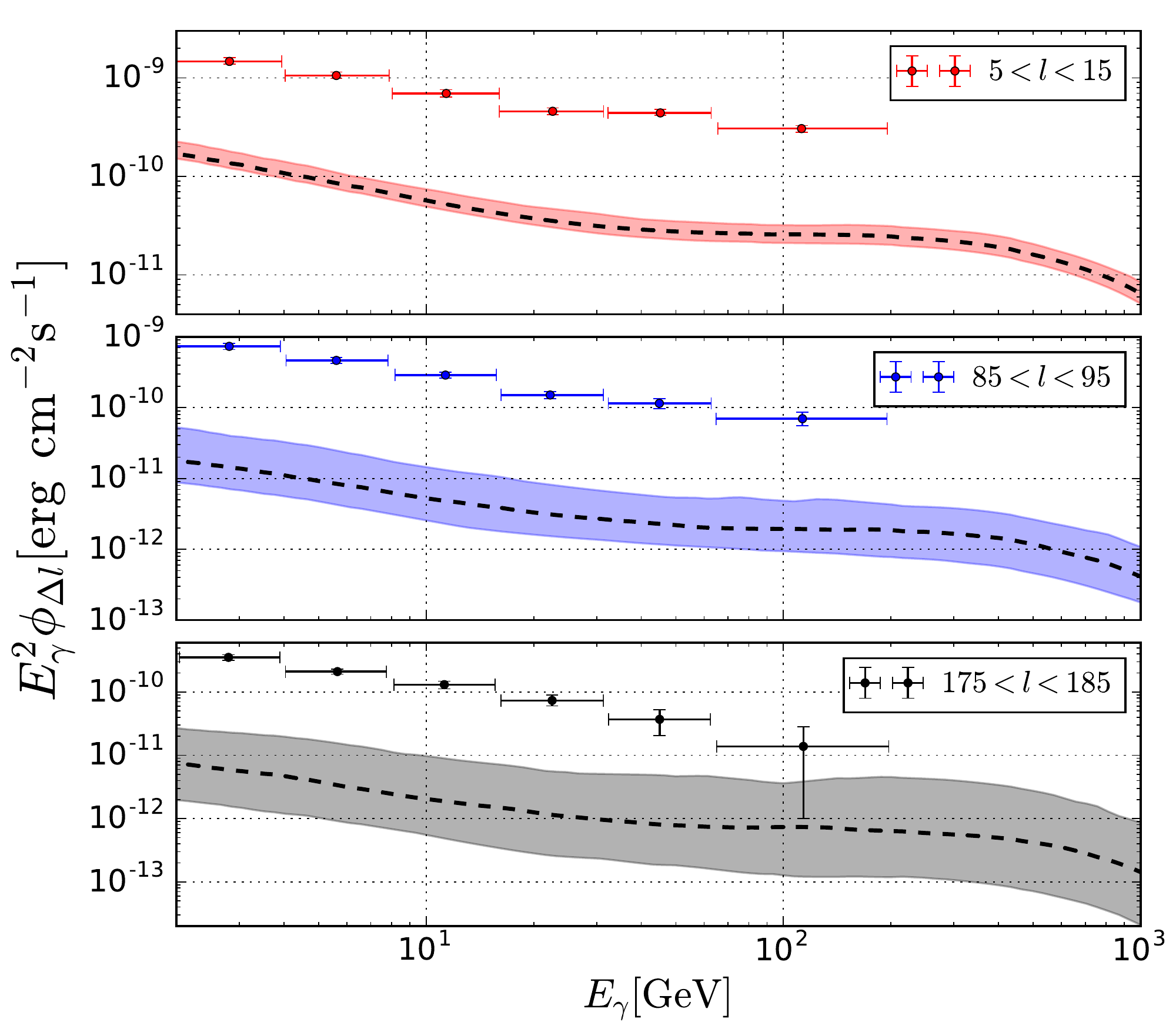} 
\includegraphics[width=1.0\columnwidth]{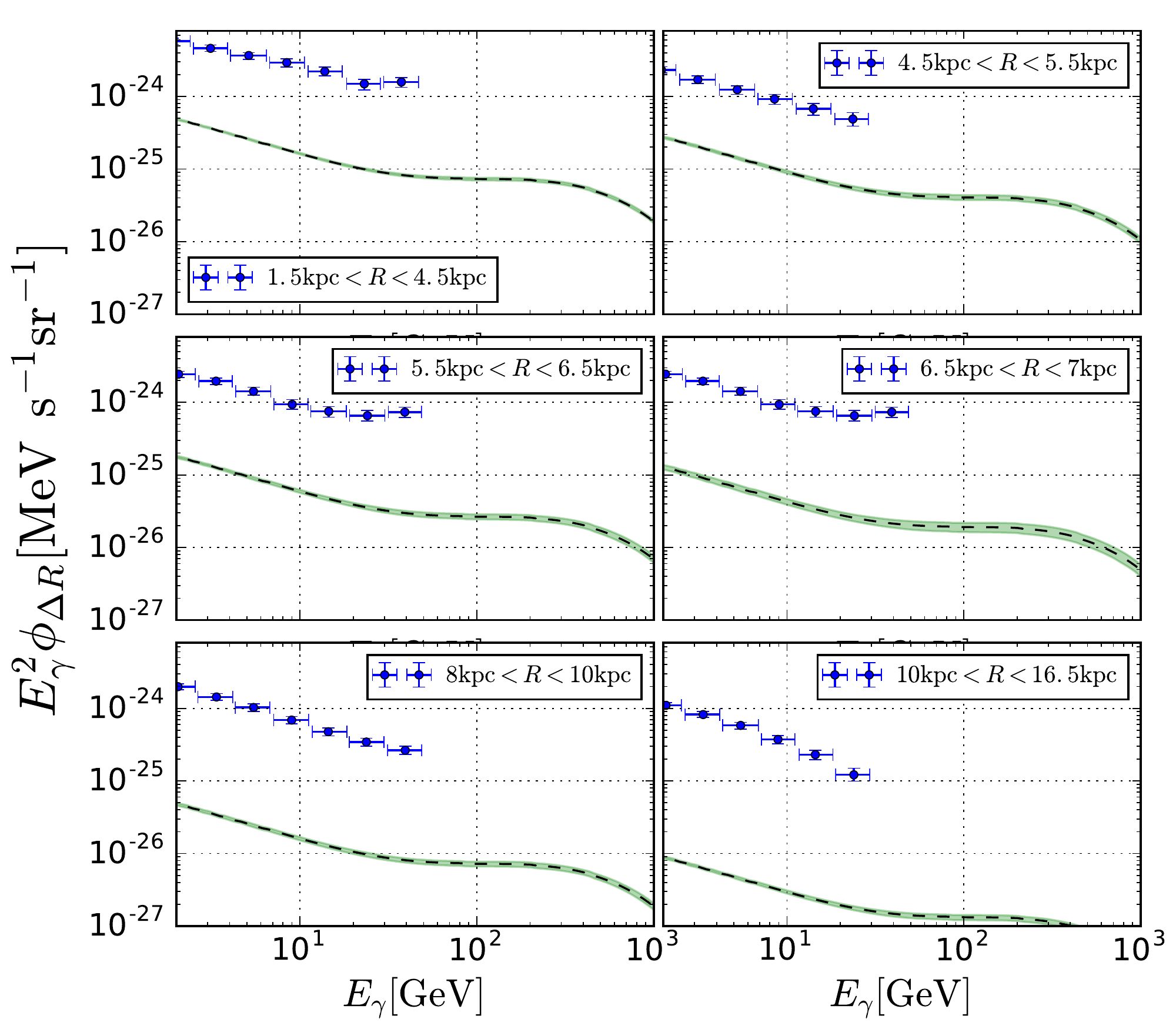} 
\caption{Same as Fig.~\ref{fig:ni045nn0}, but for case (2): $n_i = 0.45 \, \mathrm{cm}^{-3}$ and $n_n= 0.05 \, \mathrm{cm}^{-3}$.}
\label{fig:ni045nn005}
\end{figure*}

This is immediately clear looking at Fig.~\ref{fig:ni045nn005}, where we show our results for the case of partially ionized media with ion density $n_{i}=0.45 \mathrm{cm}^{-3}$ and $n_n= 0.05 \mathrm{cm}^{-3}$. IND drastically reduces the CR halo contribution to the diffuse $\gamma$-ray emission.
What is interesting to notice is that at $E_{\gamma} \gtrsim 10^2 \, \mathrm{GeV}$, which corresponds to CRs with $E \gtrsim 10^3 $ GeV, there is a change of slope in both the $\gamma$-ray flux and the $\gamma$-ray emissivity per H atom. This is due to the fact that IND is less efficient in damping long wavelength turbulence that resonate with $E \gtrsim 10^3 $ GeV particles, as can be seen from the escape time plotted in Fig.~\ref{fig:tesc}.

Clearly, the case in which the gas around the explosion is totally ionized and very rarefied, namely $n_i=0.01$ cm$^{-3}$, though interesting from the point of view of confinement time, leads to no appreciable contribution to the diffuse gamma ray emission because of the lack of target.

\begin{figure*}
\centering
\includegraphics[width=1.0\columnwidth]{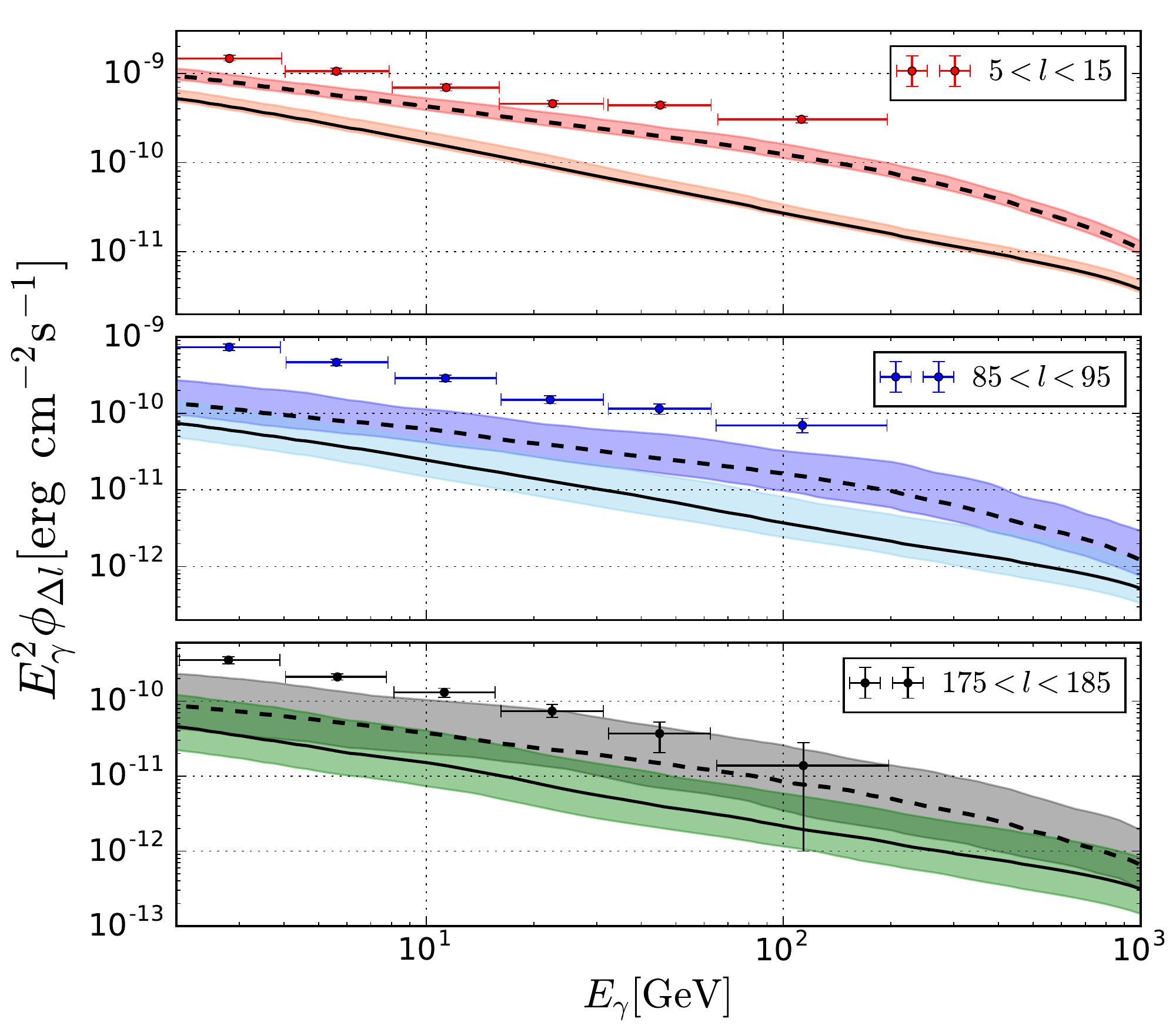} 
\caption{Comparison of the model with the data by \citep{yang2016} in the case where a steeper injection spectrum is assumed for the accelerated particles. Fully ionized ISM is considered with $n_i = 0.45 \, \mathrm{cm}^{-3}$. For all the three panels the black dashed line indicates the mean value for $\alpha = 4$ while the dotted black line indicates the one for $\alpha = 4.2$, while the shaded bands around each line represent the 68\% confidence level.}
\label{fig:comp_slope_ni_045_nn_0}
\end{figure*}

Finally, we note that when the injection slope is $\alpha = 4.2$, the confinement time is significantly reduced at high energies and as a consequence also the predicted  $\gamma$-ray emission produced by confined CRs becomes lower, as can be seen both from Fig.~\ref{fig:comp_slope_ni_045_nn_0} for the case with no neutrals and $n_i = 0.45 \, \mathrm{cm}^{-3}$. 

\section{Discussion and conclusions}
\label{sec:conclude}

Non-linear effects excited by CRs escaping their parent source may induce a considerable enhancement of the diffusion time over a spatial scale comparable with the coherence scale of the local Galactic magnetic field. In turn, since the density of the gas in the near source region is expected to be of the order of the mean Galactic disc density, $\sim 1~\rm cm^{-3}$, the grammage accumulated by CRs while escaping this region can be non negligible \cite[]{dangelo2016}. Here we discussed the implications of this scenario in terms of gamma ray production from diffuse halos around sources. While the emission from a single halo is too faint to be detected with current gamma-ray telescopes, the integration of the gamma ray emission from many halos along a given line of sight may contribute to the diffuse emission of the Galaxy in the gamma ray band.

The signal is the strongest when neutral hydrogen is absent because when this occurs, ion-neutral damping can be ignored and self-generated waves can grow to much higher levels. In this case the resulting diffuse gamma ray emission in a broad region around the central part of the Galaxy can be saturated by the emission of the extended halos around sources. Fluctuations are expected in the diffuse emission if a large contribution to the integral along the line of sight comes from these halos, since their number will be subject to Poisson statistics.
Such fluctuations become more severe along lines of sight that point towards the external part of the Galaxy, where a smaller number of sources is expected. In such directions, the contribution of near source halos to the  diffuse gamma ray background is somewhat reduced. 

In this article we have analysed one of the observational implications (the most straightforward one) of a scenario in which CRs accumulate an appreciable fraction of the total grammage in the near source region. Whether such a scenario is physically plausible and compatible with all the observational data is an important question, especially after the appearance of some alternative and rather unorthodox models of CR transport, such as those recently discussed by \cite{2010PhRvD..82b3009C} and \cite{2017PhRvD..95f3009L}. These papers speculate that a picture in which grammage is accumulated locally might in fact explain the positron excess and perhaps the seemingly anomalous behaviour of the spectrum of antiprotons. The radical change of  paradigm that these models would entail forces us to critically consider what the possible conditions are, if any, for their assumptions to be plausible. In the context of the non-linear picture proposed by \cite{dangelo2016} and investigated further in the present article, this question is equivalent to understanding the reliability of the prediction that waves can be generated effectively by CRs escaping their sources. While the basic physical processes involved in the self-generation are well understood, the relative importance of different branches of the streaming instability induced by CRs is less clear. Here we limited our attention to growth (and damping) of Alfv\'enic resonant modes, but a non-resonant mode \cite[]{2004MNRAS.353..550B} is known to grow faster provided the current of escaping CRs is large enough. If particles are streaming at the speed of light this condition translates to requiring that the energy density of escaping particles is larger than the energy density of the background magnetic field. It is easy to understand that this condition is certainly satisfied at some distances from the SNR (or some times after the explosion) but it is by no means easy to describe it quantitatively. In fact, the diffusion coefficient, that for resonant modes is easy to associate with particles present at a given location, is hard to calculate for non-resonant modes. The reason for this is that waves are initially excited on scales much shorter than the resonant scale of particles dominating the current and eventually reach a larger scale through inverse cascade only at later times.

A realistic description of the role of non-resonant modes requires investigation tools that go way beyond the ones discussed by \cite{dangelo2016} and that perhaps do not exist yet. On the other hand, it is safe to say that the excitation of non-resonant modes can only make confinement of CRs close to the sources more effective, hence the predictions discussed here should be considered as conservative. 

There is another aspect of the problem of CR self-confinement that deserves further discussion: the presence of even small amounts of neutral hydrogen (few percent) in the near source regions makes the impact of ion-neutral damping on the growth of resonant modes rather strong and reduces drastically the grammage that CRs can traverse in the near source regions. In this sense, the contribution estimated in this article to the diffuse gamma ray emission should be considered as an upper limit in that it is perhaps unrealistic to imagine that all SNe explode in dense ionized regions. On the other hand, there are at least a couple of considerations worth making when considering this issue. First, when non-resonant modes get excited their growth is typically very fast and the waves are, at least  in the beginning, almost purely growing (non propagating). Hence the role of ion-neutral damping is expected to be much less of a concern for these waves. Indeed the reduced effectiveness of ion-neutral damping on the growth of non-resonant waves has been shown quantitatively (although in a rather different physical scenario, and hence for a different set of parameters than the ones relevant here) by \cite{Reville2007}. Second, in the calculations discussed so far, the assumption was made that the medium outside a SNR has homogeneous properties over a scale of $\sim 100-200$ pc. This is also unlikely to be so, in that the environment around SNRs is often populated by dense molecular clouds. Their impact in terms of accumulated grammage may be very important: for instance if the spatial scale of the cloud is $R_{c}$ and the mean cloud density is $n_{c}$, then, even if the propagation of CRs inside the cloud were ballistic (no diffusion) the contribution of the cloud to the grammage from an individual region of size $\lambda$ around a source would dominate over the diffuse contribution if $n_{c}\left(\frac{R_{c}}{\lambda}\right)>n_{i}$. For typical parameters of the problem, $R_{c}\sim 10$ pc, $n_{c}\sim 200~\rm cm^{-3}$ and $\lambda=100$ pc, the cloud's contribution is dominant for any reasonable value of $n_{i}$, so that even if only a fraction of SNRs had a molecular cloud nearby, the total grammage contributed by near source regions may still be comparable with the total grammage as currently attributed to CR transport in the Galaxy. 

Both the issues just illustrated are currently being investigated further, but it is clear that, as discussed in the present article, CR confinement in the near source regions should first be sought after in terms of signatures in the gamma ray sky where we have the proper tools to find important evidence. In this sense, better analyses of the spatial distribution of the diffuse gamma ray emission as a function of energy, and especially of its fluctuations along different lines of sight, as discussed in this article, might be good indicators of the role of CR induced non-linear effects in the near source regions.

\section*{Acknowledgments}
The authors are very grateful to the members of the Arcetri and GSSI High Energy Astrophysics groups for numerous discussions on the topic and for continuous, fruitful collaboration on related problems.



\bibliographystyle{mnras}
\bibliography{References} 





\bsp	
\label{lastpage}
\end{document}